\magnification=\magstep1
\input epsf
\overfullrule=0pt
\centerline{\bf DECOHERENCE, CHAOS, QUANTUM-CLASSICAL CORRESPONDENCE,}
\centerline{\bf AND}
\centerline{\bf THE ALGORITHMIC ARROW OF TIME}
\bigskip
\centerline{Wojciech H. Zurek}
\medskip
\centerline{Theoretical Astrophysics}
\centerline{T-6, Mail Stop B288, LANL}
\centerline{Los Alamos, New Mexico 87545}
\bigskip
\centerline {\bf Abstract}
\medskip
\noindent The environment -- external or internal degrees of freedom coupled
to the system -- can, in effect, monitor some of its observables. As a result, 
the eigenstates of these observables decohere and behave like classical states: 
Continuous destruction of superpositions leads to environment-induced 
superselection (einselection).  Here I investigate it in the context of quantum 
chaos (i. e., quantum dynamics of systems which are classically chaotic). 
I show that the evolution of a chaotic macroscopic (but, ultimately, quantum)
system is not just difficult to predict (requiring accuracy exponentially 
increasing with time) but quickly ceases to be deterministic in principle as  
a result of the Heisenberg indeterminacy (which limits the resolution available 
in the initial conditions). This happens after a time $t_{\hbar}$ which is only 
logarithmic in the Planck constant. A definitely macroscopic, if somewhat
outrageous example is afforded by various components of the solar system which
are chaotic, with the Lyapunov timescales ranging from a bit more then a month 
(Hyperion) to millions of years (planetary system as a whole). On the timescale 
$t_{\hbar}$ the initial minimum uncertainty wavepackets corresponding 
to celestial bodies would be smeared over distances of the order of radii of 
their orbits into ``Schr\"odinger cat - like'' states, and the concept of  
a trajectory would cease to apply. In reality, such paradoxical states are 
eliminated by decoherence which helps restore quantum-classical correspondence. 
The price for the recovery of classicality is the loss of predictability:  
In the classical limit (associated with effective decoherence, and not just
with the smallness of $\hbar$) the rate of increase of the von Neumann
entropy of the decohering system is independent of the strength of the coupling 
to the environment, and equal to the sum of the positive Lyapunov exponents. 
Algorithmic aspects of entropy production are briefly explored to illustrate 
the effect of decoherence from the point of view of the observer. We show that
``decoherence strikes twice'', introducing unpredictability into the system
and extracting quantum coherence from the observer's memory, where it enters
as a price for the classicality of his records. 
\vfill
\eject
\noindent{\bf 1. Introduction}

Movements of planets have served as a paradigm of order and predictability 
since ancient times. This view was not seriously questioned until the time of
Poincar\'e, who has initiated the enquiry into the stability of the solar 
system$^1$ and thus laid foundations of the subject of dynamical chaos.
However, only recently and as a result of sophisticated numerical experiments
are the questions originally posed by Poincar\'e being answered. Two groups, 
using very different numerical approaches, have reported that the solar system 
is chaotically unstable$^{2,3}$. The characteristic {\it Lyapunov exponent} 
which determines the rate of divergence of neighboring trajectories in the phase
space is estimated to be $\lambda = (4 \times 10^6)^{-1}~{\rm [year^{-1}]}$. 
Fortunately (and in accord with the overwhelming experimental evidence) it is 
likely that this instability will not alter crucial characteristics of 
the orbits of planets such as their average distance from the sun 
(although eccentricities of the orbits may not be equally safe$^3$). 
Rather, it is the location of the planet along its orbit which is exponentially
susceptible to minute perturbations. Even trajectories of the massive outer planets alone 
appear to be exponentially unstable, although the Lyapunov exponent for that
subsystem of the solar system is harder to estimate$^2$, and may correspond to
a timescale as short as few million years, or as long as 30 Myr. 

While the instability of the planetary system takes place on a relatively long 
timescale, there are celestial bodies which become chaotically unpredictable
much more rapidly. Perhaps the best studied example is Hyperion, one of the 
moons of Saturn. Hyperion is shaped as an elongated ellipsoid. 
The interaction between its quadrupole moment and the gravitational field of 
Saturn leads to chaotic tumbling, which results in an exponential divergence 
on a timescale approximately equal to twice its 21 day orbital period$^4$. 
There are also numerous examples of chaos in the asteroid belt (such as Chiron) 
with exponential instability timescales of few hundred thousand years$^5$. 

In spite of its obviously macroscopic characteristics the solar system is, 
ultimately, undeniably {\it quantum}. This is simply because its constituents
are subject to quantum laws. The action associated with the solar system is,
of course, enormous:
$$ I \ \simeq \ {{G M_{\odot} M_J} \over R_J} \times \tau_J \simeq 1.2 \times
 10^{51} {\rm [erg~s]} \ , \eqno(1.1) $$
where the mass $M_J$ and period $\tau_J$ of Jupiter were used in the estimate.
Given this order of magnitude of $I$ and the smallness of the Planck constant
($\hbar = 1.055 \times 10^{-27}$~erg~s), one might have anticipated that the
dynamics of the solar system is a safe distance away from the quantum regime. 
However, and as a consequence of the chaotic character of its evolution, this is
{\it not} the case. 

I will begin by showing that the macroscopic size of a system -- any system --
does not suffice to guarantee its classicality. Thus, quantum theory 
implies that even the solar system
-- and every other chaotic system -- is {\it in principle} indeterministic, and 
not just ``deterministically chaotic'': Classical predictability in the chaotic 
context would require an ever increasing accuracy of its initial conditions.
This is possible in principle in classical physics. But, according to quantum 
mechanics, simultaneously increasing the accuracy of position and momentum would
eventually violate the Heisenberg principle. This time defines the quantum
predictability horizon. It is surprisingly short, and -- at least for some 
of its components -- definitely less than the age of the solar system. 

Classicality is restored with the help of environment-induced decoherence,
which continuously destroys the purity of the wavepackets. The resulting loss of
predictability can be quantified through the rate of entropy production. For a 
decohering chaotic system we shall see that this rate is; (i) independent of the
strength of the coupling to the environment, and (ii) given by sum of the
positive Lyapunov exponents. That is, the {\it quantum} entropy production rate 
coincides with the Kolmogorov-Sinai entropy in open systems, even though its 
ultimate cause is the loss of the information to the correlations with
the environment. 

The observer's own point of view of this process is briefly explored, and the 
relation between the algorithmic randomness of the observer's records and the 
von Neumann entropy of the evolving system is noted. We point out that when 
the observer is monitoring a decohering quantum system, entropy can increase as 
a result of the information loss to the environment caused by; (a) ``reduction 
of the state vector'', i. e., decoherence of the observer's record states, and;
(b) through the more usual channel -- decoherence in the monitored system. 
\bigskip

\noindent{\bf 2. Quantum predictability horizon: How the correspondence is lost.}

As a result of chaotic evolution, a patch in the phase space which 
corresponds to some regular (and classically ``reasonable'') initial condition 
becomes drastically deformed: Classical chaotic dynamics is characterized by 
the exponential divergence of trajectories. Moreover, conservation of the volume
in the phase space in the course of Hamiltonian evolution (which is initially
a good approximation for sufficiently regular initial conditions even in
cases which are ultimately quantum) implies that 
the exponential divergence in some of the directions must be balanced 
by the exponential squeezing -- convergence of trajectories -- in the other 
directions. It is that squeezing which forces a chaotic system to explore the
quantum regime: As the wavepacket becomes narrow in the direction corresponding 
to momentum;
$$ \Delta p (t) \ = \ \Delta p_0 ~ \exp (- \lambda t)  \eqno(2.1)$$
(where $\Delta p_0$ is its initial extent in momentum, and $\lambda$ is the 
relevant Lyapunov exponent) the position becomes delocalized: The wavepacket 
becomes coherent over the distance $\ell(t)$ which can be inferred from 
Heisenberg's principle:
$$ \ell (t) \geq (\hbar / \Delta p_0) \exp (\lambda t) \ . \eqno(2.2)$$
Coherent spreading of the wavepacket over large domains of space is disturbing
in its own right. Moreover, it may lead to a breakdown of the correspondence 
principle at an even more serious level: Predictions of the classical 
and quantum dynamics concerning some of the expectation values no longer
coincide after a time $t_{\hbar}$ when $\ell (t)$ reaches the scale on which
the potential is nonlinear.

Such a scale $\chi$ can usually be defined by comparing the classical force 
(given by the gradient of the potential $\partial_x V$) 
with the leading order nonlinear contribution $\sim \partial^3_x V$:
$$\chi \ \simeq \ \sqrt {{\partial_x V} \over {\partial^3_x V}} \ . \eqno(2.3)$$
For the gravitational potential $\chi \simeq R / \sqrt{2}$, where $R$ is 
a size of the system (i. e., a size of the orbit of the planet). The reason
for the breakdown of the correspondence is that when the coherence length
of the wavepacket reaches the scale of the nonlinearity, 
$$ \ell (t) \ \simeq \chi \ , \eqno(2.4)$$
the effect of the potential energy on the motion can be no longer represented 
by the classical expression for the force$^8$, $F(x) = \partial_x V(x)$, since 
it is not even clear where the gradient is to be evaluated for a delocalized 
wavepacket. As a consequence, after a time given by:
$$ t_{\hbar} \ = \ \lambda^{-1} \ln {{\Delta p_0 \chi} \over \hbar} \ ,
\eqno(2.5)$$
the expectation value of some of the observables of the system may even begin 
to exhibit noticeable deviations from the classical evolution.

This is also close to the time beyond which the combination of classical 
chaos and Heisenberg's indeterminacy makes it impossible {\it in principle} 
to employ the concept of a trajectory. Over the time $\sim t_{\hbar}$ a chaotic 
system will spread from a regular Planck-sized volume in the phase space into 
a (possibly quite complicated) wavepacket with the dimensions of its envelope
comparable to the range of the system. This timescale defines the quantum 
predictability horizon -- a time beyond which the combination of classical
chaos and quantum indeterminacy makes predictions not just exponentially 
difficult, but impossible in principle. The shift of the origin of the loss 
of predictability from classical deterministic chaos to quantum indeterminacy   
amplified by the exponential instabilities is just one of the symptoms of the
inability of classical evolutions to track the underlying quantum dynamics.\footnote*{One may be concerned that the argument sketched out above may be inconsistent 
-- that the requirements of {\it quantum} Heisenberg's indeterminacy cannot be 
applied to systems which are {\it classical}. Below, I shall carry out a completely 
quantum analysis of this issue. But I believe that the above sketch of an 
argument is essentially correct. For, even ostensibly classical systems must 
respect the Heisenberg's principle: The Einstein-Bohr double slit experiment
debate was settled, after all, with an appeal to the constraints imposed by the 
Heisenberg's principle on the accuracy of simultaneous measurements of position
and momentum of the movable slit introduced by Einstein!}

This breakdown of correspondence can be investigated more rigorously by 
following the evolution generated for the possibly macroscopic, yet ultimately
quantum system by the {\it Moyal bracket} (that is, a Wigner transform of the 
quantum von Neumann equation). The Moyal bracket can be expressed through 
the familiar classical Poisson bracket as:
$$ \{H,W\}_{MB} \ = \ -i \sin(i \hbar \{H,W\}_{PB})/\hbar \ . \eqno(2.6)$$
Above, $H$ is the Hamiltonian of the system, and $W$ is an object in the phase 
space (i. e., a probability distribution). In our quantum case, $W$ will denote
a Wigner function -- a Wigner transform of the density matrix. 

When the potential $V$ in $H$ is analytic, the Moyal bracket can be expanded 
in powers of the Planck constant. Consequently, the evolution of $W$ is given by:
$$ \dot W \ = \ \{H,W\}_{PB} ~ + 
~ \sum_{n\geq 1}{{\hbar^{2n} (-)^n} \over
{2^{2n} (2n+1)!}} \partial_x^{2n+1} V (x) \partial_p^{2n+1} W (x,p) \ . \eqno(2.7)$$
Correction terms above will be negligible when $W(x,p)$ is a reasonably
smooth function of $p$, that is when the higher derivatives of $W$ with
respect to momentum are small. However, the Poisson bracket alone predicts that,
in the chaotic system, they will increase exponentially quickly as a result of 
the ``squeezing'' of $W$ in momentum, Eq. (2.1). Hence, after $t_{\hbar}$
quantum ``corrections'' will become comparable to the first classical term on 
the right hand side of Eq. (2.7). At that point the Poisson bracket will no 
longer suffice as an approximate generator of evolution. The phase space 
distribution will be coherently extended over macroscopic distances, and 
interference between the fragments of $W$ will play a crucial role.

The timescale on which the quantum - classical correspondence is lost in 
a chaotic system can also be estimated (or, rather, bounded from above) by the
formula$^{6,7}$:
$$ t_r \ = \ \lambda^{-1} \ln (I / \hbar) \ , \eqno(2.8)$$
where $I$ is the action which -- for the solar system -- we have already 
estimated, Eq. (1.1). It follows that, for the planetary system,  
quantum-classical correspondence should be lost after approximately:
$$ t_r \ \simeq \ 711 ~ {\rm [Myr]} \ .$$
This is less than a fifth of the modest estimates of the age of Earth, and, 
presumably, a still smaller fraction of the actual age of the solar system.
When we compute instead the value of $t_{\hbar}$, setting initial uncertainty 
in the momentum to $\Delta p_0 = \hbar / \Lambda_{dB}(T) \simeq 10^9$~[g~cm/s], 
where we take $\Lambda_{dB}(T)$ for definiteness to be the thermal de Broglie wavelength of Jupiter at its 
present surface temperature of $\sim 100~K$, we estimate almost identically:
$$ t_{\hbar} \ \simeq \ 682 ~ {\rm [Myr]} \ . $$
A similar calculation for Hyperion results in a much smaller (and, 
therefore, so much more disturbing):
$$ t_{\hbar} \ \simeq \ 20 ~ {\rm [yr]} \ . $$
Moreover, it should be pointed out that -- in the macroscopic regime considered
here -- the above estimates are exceedingly {\it insensitive} to either 
the action or the initial momentum  uncertainty: Both of these quantities 
appear inside the logarithm. 

\bigskip
\noindent {\bf 3. The solar system as a Schr\"odinger cat}

We have seen above that a seemingly very secure prediction of quantum physics
as applied to the solar system fails: According to the Schr\"odinger equation, 
less than a billion years after its formation the behavior of the solar 
system  should be flagrantly non-classical, with the quantum states of 
celestial bodies spread over dimensions comparable with the sizes of 
their orbits, and with the planetary dynamics no longer 
in accord with the laws of Newton! Somehow, this does not seem to be the case. 
The source of the paradox is obvious: Chaotic dynamics increases the size of 
the coherent wavepacket with $\exp(\lambda t)$, so that it becomes comparable 
with the dimensions of the solar system after a time $t_{\hbar}$, which is but 
a fraction of its age. Similarly, and after only $\sim$ 20 years the quantum 
state of Hyperion would be a coherent superposition involving macroscopically 
distinct orientations of its major axes. 

There are parallels between
our discussion above and the famous argument due to Schr\"odinger$^{10}$ in
that a very macroscopic object (planet here, cat there) is forced, through the 
strict compliance with the laws of quantum mechanics, into a very non-local
state, never encountered as an ingredient of our familiar ``classical reality''.
The main difference between the two examples -- the Schr\"odinger cat and the 
classically chaotic but ultimately quantum planet -- is in the manner in which they are forced into the
final superposition: Schr\"odinger cat either lives (or dies) as a result of 
decay of an unstable nucleus: An intermediate step in which a quantum state of 
the nucleus is {\it measured} (to determine the fate of the cat) is essential. 
Thus, in the case of the cat it was possible to entertain the notion that the 
(admittedly preposterous) final superposition of dead and alive cat
could be avoided if the process of measurement was properly understood. This
``way out'' is no longer available in the case of celestial bodies we are
discussing. They evolve into states which are nonlocal and flagrantly
quantum simply as a result of dynamical evolution -- the measurement plays 
absolutely no role in setting up the paradox. And the systems involved are 
certainly even more macroscopic than the cat. Moreover, if the reader considers
the idea of putting a living being (cat) in a superposition especially 
tantalizing, this is certainly occurring also in the case considered here.
For, in accord with the quantum arguments presented
above, after a time $t_{\hbar}$ Earth would evolve into a state corresponding
to a coherent
superposition of all the seasons, as well as all of the hours of the day!

So what is the resolution of the above paradox? Let us
us start with a few possibilities which may be tempting at the first sight,
but which ultimately lead nowhere. To begin with, one might be worried that in 
the arguments above we have cut corners by considering just one 
spatial dimension and one momentum, while the solar system is inhabiting 
a multidimensional phase space. This is certainly true, but the squeezing 
in momentum and the resulting delocalization is unlikely to be alleviated 
by considering multidimensional phase space of all the celestial bodies for 
which it occurs. This is especially true for systems such as Hyperion or Chiron,
which have a rather large Lyapunov exponent. Another more contrived possibility
of avoiding the difficulty with quantum-classical correspondence at present
would be to design an initial state which evolves into a ``classical looking'' 
state by the present epoch. This can in principle be done, but -- as the
reader is invited to verify -- it requires initial states which are as
flagrantly quantum as those we were forced to consider above. 

Finally and in desperation one might consider abandoning quantum theory for 
some other theory which is almost exactly like quantum theory (to pass all of
the experimental tests) but contains either nonlinear corrections,
or allows for underlying ``hidden variable'' dynamics, or, 
perhaps, introduces ``collapse of the wavepacket'' {\it ad hoc} at some 
fundamental level in order to get rid of the quantum nonlocality. 
All of these ideas face either serious experimental constraints 
(which render them useless for the purpose of making quantum theory look 
classical) or profound theoretical difficulties (such as a conflict with 
the Lorentz invariance), or both.

\bigskip
\noindent {\bf 4. Decoherence, the quantum, and the classical}

I shall instead contend that quantum theory is rigorously correct, but that
the superposition principle cannot be applied naively, especially to the 
macroscopic objects. The failure of such a simple-minded application of quantum
principles to the classical domain is, however, itself a consequence of the
unitarity of quantum evolution: Macroscopic objects are all but impossible
to insulate from their environments. Consequently, external and internal 
degrees of freedom continuously ``monitor" -- that is, become correlated --
with their state$^{11-13}$. This is the process of decoherence$^{11-20}$. 

We have no room here to develop the theory of decoherence systematically and
completely. A sketch with a few leads to the existing (and rapidly
expanding) literature will have to suffice. The key point is the observation
that in quantum mechanics information matters much more than in classical
mechanics, where it can be acquired without influencing in any way the 
{\it actual} state of the system, which exists and evolves independently of
what is known about it. Such a neat division between information and ``physical
reality'' is impossible to implement in the quantum realm: Acquisition of 
information is equivalent to the establishment of a correlation, which in turn
is reflected in the loss of the capacity for interference. The double slit 
experiment is a classic example. As soon as it is known through which slit 
the photon has passed, the possibility for interference is lost.

The information transfer which accompanies decoherence has the same nature as 
that encountered in quantum measurements, or for that matter, in quantum 
computation. In either case it is useful to 
represent it with an elementary logical gate -- so-called ``controlled not'' 
or a ``{\tt c-not}'' -- which reversibly copies a single bit of information 
between two (two-state) quantum systems, known respectively as a ``control'' 
and a ``target'':
$$ (a|0>~+~b|1>)_c~ |0>_t ~ \longrightarrow ~ a|0>_c|0>_t ~ + ~ b|1>_c|1>_t \ .
\eqno(4.1)$$
In short, a quantum {\tt c-not} is an obvious generalization of a classical 
{\tt c-not} gate
(also occasionally known as an ``exclusive or'' or ``{\tt xor}''), which 
flips the state of the target bit when the control is in a state ``1'', and 
does nothing otherwise. The analogy between Eq. (4.1) and the von Neumann model
of a quantum measurement$^{21}$ is obvious: The control bit acts as a measured 
system, forcing the apparatus (target bit) to measure its state.

The state on the right hand side of Eq. (4.1) is, however, not the resolution 
of the measurement problem, but, rather, its cause: When $a=-b=1/\sqrt2$ 
it is in fact identical to the states encountered in the Bohm version of the 
Einstein-Podolsky-Rosen experiment$^{22,23}$. The correlation established is 
a quantum entanglement, with all its seemingly paradoxical consequences, which 
deny to each of the two systems involved the right to possess ``a state of
their own'' prior to 
a measurement$^{23-25}$. Decoherence converts entanglement into classical 
correlation by allowing the environment to carry out such additional 
``measurements'' on the to-be-classical system$^{11,12}$. Its consequence
is then a {\it disentanglement} -- correlations between the system and the apparatus 
weaken to their classically allowed strength.

Decoherence can then be conveniently ``caricatured'' (if not quite 
``represented'') by means of the {\tt c-not} like gates transferring information
about the to-be-classical observable to the environment. This is shown in 
Fig. 1, where a {\tt c-not} symbolic representation of the measurement
carried out by the target-apparatus on the control-system is also illustrated. 
In the process of decoherence information flows from the quantum correlations 
between the memory (of the observer or of the apparatus) with to-be-classical 
observable of the decohering entity to the correlation with the environment. 
Decoherence is a purely quantum effect -- information does not matter 
in classical dynamics. 
In the symbolism of Fig. 1 decoherence can be conveniently contrasted with 
the more familiar consequence of the coupling to the environment -- noise -- 
in which the state of the environment becomes inscribed on the observable of 
interest.

The direction of the information flow depicted in the {\tt c-not}s depends on 
the observables involved. This is an important consequence of the quantum nature
of the information transfer. The reader can verify this by re-writing the action
of the {\tt c-not} in the complementary basis $|\pm>=(|0>\pm |1>)/\sqrt2$, and 
checking that in the new basis the roles of the control and target are reversed.
This illustrates the connection between the loss of phase coherence and 
``reduction of the state vector'' -- while the environment is ``measuring'' 
a certain observable $A$, its conjugate (Fourier) complement is busy 
``measuring'' the state of the environment, and storing the information in 
the phases between the eigenstates of $A$ -- in the correlations with the
states $|+>$ and $|->$. 

In idealized examples of decoherence (i.e., in absence of the self-hamiltonian) 
a preferred observable is selected by the interaction Hamiltonian with the 
environment -- it satisfies (or, at least, approximates) the commutation relation$^{11,12}$:
$$ [H_{int}, A] \ = \ 0 \ . \eqno(4.2)$$
However, in more realistic circumstances involving dynamical evolution Eq. (4.2)
defines only the ``instantaneous'' preferred (pointer) observable. Long-term
predictability (which is a convenient and natural criterion$^{20}$ of the more 
elusive ``classicality'') is optimized by the states which are least perturbed 
by the environment in spite of the incessant rotation between the observables 
and their complements caused by the dynamics$^{20,26,27}$.

A system -- such as a harmonic oscillator or, for that matter, a chaotic quantum
system -- is then described by an effective {\it master equation}$^{28-31}$, 
which continuously transforms pure states into mixtures. The rate at which this 
happens is set in part by the coupling, but the nature of the initial state 
plays the decisive role. States which become least mixed are then 
most predictable and can be regarded as most classical. 
The high-temperature master equation for a particle
interacting with the thermal excitations of the environment composed of harmonic
oscillators is a convenient and often studied example$^{29-31}$. The reduced 
density matrix $\rho(x, x')$ of the system (obtained by tracing out the state 
of the environment) in the position representation evolves 
in this case according to:
$$\dot \rho \  = \  \overbrace{\underbrace{ -{i \over \hbar} [H, \rho]}_{\dot p = - FORCE = \nabla V}}^{von~
Neumann~eq.} - \ 
\overbrace{\underbrace{\gamma (x-x') ({\partial \over \partial x} 
\ - \  {\partial \over \partial x'})\rho}_{\dot p=-\gamma p}}^{relaxation} 
\  - \ \overbrace{\underbrace{{2 m \gamma k_B T \over \hbar^2} (x-x')^2 ~ \rho}}_{classical~phase~space}^{decoherence} \ .
\eqno (4.3)$$
Above, $H$ is the effective Hamiltonian of the system (i. e., with the potential
renormalized to recognize the influence of the environment), and $\gamma$ is the
relaxation rate. The interaction Hamiltonian was assumed to couple the coordinate
$x$ of the system with the coordinates of the environment oscillators. When
the oscillators are collectively represented by a field $\phi(q)$, the coupling 
can be taken to have a form$^{30}$:
$$ H_{int} = \varepsilon x \dot \phi(q,t) \ , \eqno(4.4) $$
in which case the effective viscosity is $\eta = 2 m \gamma = \varepsilon^2/4m$.

Arbitrary superpositions of localized wavepackets can in principle exist 
in the Hilbert space, but, as a result of environmental monitoring, they
are exceedingly unstable in practice: Continuous monitoring enforces
{\it environment -- induced superselection}$^{12}$: Only some -- relatively few 
-- of the quantum states which can exist in principle are capable of surviving
the interaction with the environment more or less intact. Which states can  
survive depends on the form of the interaction with the environment$^{11}$. 
The general rule is that the states which are localized in the monitored 
observables are most stable$^{11,12,20,26,27}$. Moreover, when the Hamiltonian 
of interaction is a function of some observable, then the environment is most 
effective at monitoring it. This singles out the preferred 
{\it pointer states}$^{11,17}$, Eq. (4.2). They usually turn out to be localized
in position, since the interactions tend to depend on the distance$^{12}$, 
$H_{int}=H_{int}(x)$, as in Eq. (4.4).

The tendency towards localization in position can be characterized by the time 
it takes for the two fragments of the wavepacket separated in space by the
distance $\delta x$ to lose quantum coherence. The {\it decoherence time}$^{14}$:
$$\tau_D(\delta x) \ = \ \gamma^{-1} (\Lambda_{dB}(T) / \delta x)^2 \eqno(4.5)$$
can be computed from the third term of Eq. (4.3). It is proportional to the 
relaxation time $ \tau_R = 1/\gamma$ (which determines the rate at which the 
system loses energy due to the interaction with the environment), but in the 
macroscopic realm it is much faster: For a one gram object at room 
temperature the ratio of the thermal de~Broglie wavelength 
$\Lambda_{dB}(T)=\hbar/\sqrt{2mk_BT}$
to the separation $\delta x = 1$~cm is
approximately $10^{-20}$. Hence, in the above example, and under the 
circumstances in which thermal excitations dominate the process of decoherence 
(an assumption which allows one to derive Eq. (4.3) and the simple expression, 
Eq. (4.5), but which does not effect the conclusion about the nearly 
instantaneous onset of decoherence for macroscopic objects) $\tau_D \simeq 10^{-40} /\gamma$. 

It follows that quantum coherence may be (and, for macroscopic objects, it is) 
lost exceedingly rapidly even when the relaxation time is very large. 
Even when the assumptions leading to the simple master equation (4.3) are not
valid, and Eq. (4.5) overestimates the decoherence rate, the conclusions about 
the relaxation being very slow in comparison with decoherence is unlikely to
change in the macroscopic realm.
Preliminary experimental indication which corroborates these theoretical 
expectations is just at hand, following beautiful microwave cavity 
experiments$^{31}$ reported here by Haroche. Further studies of various aspects 
of decoherence are likely to follow in the wake of the ion trap ``Schr\"odinger 
cat'' experiments$^{32}$, either as a byproduct of the quantum computation 
research described here by Wineland, or, more directly, as a consequence of 
the ``reservoir engineering'' proposal$^{33}$. 

For microscopic objects (such as an electron) and/or for microscopic 
separations, the estimate of $\tau_D$ may become comparable to $\tau_R$, and -- 
when the isolation from the environment is sufficient -- can be much larger 
than the characteristic dynamical timescales. The quantum nature 
of the evolution would then manifest itself unimpeded. But for Jupiter or 
Hyperion the opposite -- the {\it reversible classical limit}$^{17,20}$ 
with $\tau_D~ <<< t_{DYNAMICAL} <<< \tau_R$ -- is going to be enforced.

\bigskip
\noindent {\bf 5. Exponential instability vs. decoherence}

In a quantum chaotic system weakly coupled to the environment the process of 
decoherence briefly sketched above will compete with the tendency for coherent
delocalization, which occurs on the characteristic timescale given by the 
Lyapunov exponent $\lambda$. Exponential instability would spread the wavepacket
to the ``paradoxical'' size, while monitoring by the environment will attempt 
to limit its coherent extent by smoothing out interference fringes. 
The two processes shall reach {\it status quo} when their rates are comparable:
$$ \tau_D (\delta x)~ \lambda \ \simeq \ 1 . \eqno(5.1)$$
As the decoherence rate depends on $\delta x$, this equation can be solved 
for the critical, steady state coherence length, which yields $\ell_c \sim \Lambda_{dB}(T) \times \sqrt{\lambda/\gamma}$. 

A more careful analysis can be based on the combination of the Moyal bracket 
and the master equation approach to decoherence we have just sketched. 
In many cases (including the situation of large bodies immersed in the typical
environment of photons, rarefied gases, etc.) an effective approximate equation 
can be derived and translated into the phase space by performing a Wigner 
transform on Eq. (4.3). Then:
$$ \dot W \ = \ \{H,W\}_{PB}+2\gamma\partial_p pW+D \partial_p^2 W ~ + ~ 
\sum_{n\geq 1}{{\hbar^{2n} (-)^n} \over {2^{2n} (2n+1)!}} 
\partial_x^{2n+1} V (x) \partial_p^{2n+1} W (x,p) \ . \eqno(5.2)$$
The second term causes relaxation, and, in the macroscopic limit, it can be made
very small without decreasing the effect of decoherence caused by the third, 
diffusive term. Its role is to destroy quantum coherence of the fragments of 
the wavefunction between spatially separated regions. Thus, in effect, this 
{\it decoherence term} can assure that the Poisson bracket is always accurate: 
Diffusion prevents the wavepacket from becoming too finely structured in 
momentum, which -- as we have seen early on in the paper, would have caused 
the failure of the correspondence principle. In case of the thermal environment 
the diffusion coefficient $D= \eta k_B T$, where $\eta$ is viscosity. 
The competition between the squeezing due to the chaotic instability and 
spreading due to diffusion leads to a standoff when the Wigner function becomes
coherently spread over:
$$ \ell_c \ = \ \hbar \sqrt {\lambda \over {2D}} \ = \ \Lambda_{dB}(T) \times \sqrt{\lambda / 2 \gamma} \ , \eqno(5.3)$$
This translates into the critical (spatial) momentum scale of:
$$ \sigma_c \ = \ \sqrt{{2 D} \over \lambda}\ . \eqno(5.4)$$
which nearly coincides with the quick estimate, Eq. (5.1).

Returning to our outrageous example, for a planet of the size of Jupiter a 
chaotic instability on the four million year timescale and the consequent delocalization would be 
easily halted even by a very rarefied medium (0.1 atoms/cm$^3$, comparable to
the density of interplanetary gas in the vicinity of massive outer planets) at 
a temperature of 100K (comparable to their surface temperature): The resulting
$\ell_c$ is of the order of $10^{-29}$~cm! Thus, decoherence is exceedingly 
effective in preventing the packet from spreading; $\ell_c <<< \chi$, by 
an enormous margin. Hence, the paradox we have described in the first part of
the paper has no chance of materializing.

The example of quantum chaos in the solar system is a dramatic illustration of
the effectiveness of decoherence, but its consequences are, obviously, not 
restricted to celestial bodies: Schr\"odinger cats, Wigners friends, and, 
generally, all of the systems which are in principle quantum but sufficiently 
macroscopic will be forced to behave in accord with classical mechanics as a 
result of the environment - induced superselection$^{11,12}$. This will be the 
case whenever:
$$ \ell_c \ll \chi \ , \eqno(5.5)$$
since $\ell_c$ is a measure of the resolution of ``measurements'' carried 
out by the environment. 

This incredible efficiency of the environment in monitoring (and, therefore,
localizing) states of quantum objects is actually not all that surprising.
We know (through direct experience) that photons are capable of maintaining 
an excellent record of the location of Jupiter (or any other macroscopic body). 
This must be the case, since we obtain our visual information about 
the Universe by intercepting a minute fraction of the reflected (or emitted) 
radiation with our eyes. 

Our discussion extends and complements development which goes back more than 
a decade$^{35}$. We have established a simple criterion for the 
recovery of the correspondence, Eq. (5.5), which is generously met in the 
macroscopic examples discussed above. And, above all, we have demonstrated that
the {\it very same} process of decoherence which delivers ``pointer basis'' 
in the measuring apparatus can guard against violation of the quantum-classical 
correspondence in dynamics.

\bigskip
\noindent{\bf 6. The arrow of time: a price of classicality?}

Decoherence is caused by the continuous measurement-like interactions between 
the system and the environment. Measurements involve transfer  of information, 
and decoherence is no exception: The state of the environment acquires 
information about the system.  For an observer who has measured the state of 
the system at some initial instant the information he will still have 
at some later time will be  influenced (and, in general, diminished) by 
the subsequent interaction between  the system and the environment. 
When the observer and the environment monitor the same set of observables, 
information losses will be minimized. This is in fact the idea behind the 
{\it predictability sieve}$^{20}$ -- an information-based tool which allows 
one to look for the einselected, effectively classical states under quite
general circumstances. When, however, the state implied by 
the information acquired by the observer either differs right away
from the preferred basis selected by the environment, or -- as will be the case
here -- evolves dynamically into such a ``discordant'' state, environment
will proceed to measure it in the preferred basis, and, from the observers
point of view, information loss will ensue.

This information loss can be analyzed in several ways. The simplest is to 
compute the (von Neumann) entropy increase in the system. This will be our 
objective in this section. However, it is enlightening to complement this 
``external'' view by looking at the consequences of decoherence from the point 
of view of the observer, who is repeatedly monitoring the system and updating 
his records. We shall sketch such an approach (which utilizes algorithmic
information content as a measure of randomness) in Section 7.

The loss of information can be quantified by the increase of 
the von Neumann entropy:
$$ {\cal H} \ = \  -Tr~ \rho \ln \rho \eqno(6.1)$$
where $\rho$ is the reduced density matrix of the system. 
We shall now focus on the rate of increase of the von Neumann entropy in 
a dynamically evolving system subject to decoherence. As we have seen before, 
decoherence restricts the spatial extent of the quantum coherent patches to the 
critical coherence length $\ell_c$, Eq.~(5.3). A coherent wavepacket which 
overlaps a region larger than $\ell_c$ will decohere rapidly, on a timescale 
$\tau_D$ shorter than the one associated with the classical predictability loss 
rate given by the Lyapunov exponent $\lambda$. Such a wavepacket 
will deteriorate into a mixture of states each of which is coherent over
a domain of size $\ell_c$ by $\sigma_c=\hbar/\ell_c$. Consequently, the
density matrix can be approximated by an incoherent sum of reasonably localized
and approximately pure states. When $N$ such states contribute more or less 
equally to the density matrix, the resulting entropy is ${\cal H}  \ \simeq  \ \ln N.$

The coherence length $\ell_c$ determines the resolution with which the 
environment is monitoring the state of a chaotic quantum system. That is, 
by making an appropriate
measurement on the environment one could in principle localize the system to 
within $\ell_c$. As time goes on, the initial phase space patch 
characterizing the observer's information about the state of the system will be 
smeared over an exponentially increasing range of the coordinate, Eq. (2.2).
When the evolution is reversible, such stretching does not matter, at least
in principle: It is matched by the squeezing of the probability density in 
the complementary directions (corresponding to negative Lyapunov exponents). 
Moreover, in the quantum case folding will result in the interference -- 
a telltale signature of the long range quantum coherence, best visible in the
structure of the Wigner functions. 

Narrow wavepackets, and, especially, small-scale interference fringes are 
exceedingly susceptible to the monitoring by the environment. Thus, the
situation changes dramatically 
as a result of decoherence: 
In a chaotic quantum system the number of independent eigenstates of 
the density matrix will increase as:
$$ N \ \simeq \ \ell(t)/\ell_c \ \simeq \ {\hbar \over { \Delta p_0 \ell_c}} ~
\exp(\lambda t) \eqno(6.2)$$
Consequently, the von Neumann entropy will grow at the rate:
$$ \dot {\cal H} \simeq \ {d \over {dt}} \ln (\ell(t)/\ell_c)  \ \simeq \ \lambda
\eqno(6.3)$$
This is a ``corollary'' of our discussion, and perhaps even its key result:
Decoherence will help restore the quantum-classical correspondence. But we have
now seen that this will happen at a price: Loss of information is an inevitable 
consequence of the eradication of the ``Schr\"odinger cat'' states which were
otherwise induced by the chaotic dynamics. They disappear because 
the environment is ``keeping an eye'' on the phase space, monitoring the 
location of the system with an accuracy set by $\ell_c$.

Throughout this paper we have ``saved'' on notation, using ``$\lambda$'' 
to denote (somewhat vaguely) the rate of divergence of the trajectories 
of the hypothetical chaotic system. It is now useful to become a bit more 
precise. A Hamiltonian system with ${\cal D}$ degrees of freedom will have in 
general many (${\cal D}$) pairs of Lyapunov exponents with the same absolute value but 
with the opposite signs. These global Lyapunov exponents are obtained by averaging
local Lyapunov exponents, which are the eigenvalues of the local transformation,
and which describe the rates at which a small patch centered on a trajectory 
passing through a certain location in the phase space is being deformed. The 
averaging of the local exponents is achieved by following 
the trajectory of the system for a sufficiently long time.

The evolution of the Wigner function in the phase space is governed by the 
local dynamics. However, over the long haul, and in the macroscopic case, the
patch which supports the probability density of the system will be exponentially
stretched. The stretching and folding will produce a phase space structure which
differs from the classical probability distribution because of the presence of 
the interference fringes, with the fine structure on the (momentum) scale 
of the order $\hbar/\ell^{(i)}_c(t)$. In an isolated system
this fine structure will saturate only when the envelope of the Wigner function 
will fill in the available phase space volume. Monitoring by the environment
destroys these small scale interference fringes and keeps $W$ from becoming 
narrower than $\sigma_c$ in momentum. As a result -- and in 
accord with Eq.~(6.3) above -- the entropy production will asymptotically 
approach the rate given by the sum of the positive Lyapunov exponents:
$$ \dot {\cal H} \ = \ \sum_{i=1}^{\cal D} \lambda^{(i)}_+  \ . \eqno(6.4)$$
This result$^8$ is at the same time familiar and quite surprising. It is 
familiar because it coincides with the Kolmogorov-Sinai formula for 
the entropy production rate for a {\it classical} chaotic system. Here we have 
seen underpinnings of its quantum counterpart. It is
surprising because it is independent of the strength of the coupling between the
system and the environment, even though the process of decoherence (caused by 
the coupling to the environment) is the ultimate source of entropy increase.

This independence is indeed remarkable, and leads one to suspect that the
cause of the arrow of time may be traced to the same phenomena which are
responsible for the emergence of classicality in chaotic dynamics, and elsewhere
(i. e., in quantum measurements). In a sense, this is of course not a complete
surprise: Von Neumann knew that measurements are irreversible$^{21}$. 
And Zeh$^{36}$ emphasized the close kinship between the irreversibility of
the ``collapse'' in quantum measurements, and in the second law, and cautioned 
against circularity of using one to solve the other. What is however surprising
is both that the classical-looking result has ultimately quantum roots, and that
these roots are so well hidden from view that the entropy production rate 
depends solely on the classical Lyapunov exponents.

The environment may not enter explicitly into the entropy production rate, 
Eq. (6.4), but it will help determine when this asymptotic formula becomes
valid. The Lyapunov exponents will ``kick in'' as the dimensions of the patch
begin to exceed the critical sizes in the corresponding directions,
$\ell^{(i)}(t)/\ell_c^{(i)} > 1  $.
The instant when that happens will be set by 
the strength of the interaction with the environment, which determines $\ell_c$.
This ``border territory'' may be ultimately the 
best place to test the transition from quantum to classical. One may, for 
example, imagine a situation where the above inequality is comfortably satisfied
in some directions in the phase space, but not in the others. In that case 
the rate of the entropy production will be lowered to include only these
Lyapunov exponents for which decoherence is effective.
\bigskip

\noindent{\bf 7. Decoherence, einselection, and the observer: Algorithmic view
of entropy production.}

The significance of the efficiency of decoherence goes beyond the example of the
solar system or the task of reconciling quantum and classical predictions
for classically chaotic systems. Every degree of freedom coupled to the
environment will suffer loss of quantum coherence. Objects which are more 
macroscopic are generally more susceptible. In particular, the ``hardware''
responsible for our perceptions of the external Universe and for keeping 
records of the information acquired in course of our observations is obviously
very susceptible to decoherence: Neurons are strongly coupled to the 
environment, and are definitely macroscopic enough to behave in 
an effectively classical fashion. That is, they have a
decoherence timescale many orders of magnitude smaller than the relatively 
sluggish timescale on which they can exchange and process information. 
As a result, in spite of the undeniably quantum nature of the fundamental 
physics involved, perception and memory have to rely on the information stored
in the decohered (and, therefore, effectively classical) degrees of freedom. 

An excellent illustration of the constraint imposed on the information 
processing by decoherence comes from the recent discussions of the possibility
of implementation of quantum computers: Decoherence is viewed as perhaps 
the most serious threat to the ability of a quantum information processing 
system to carry out a superposition of computations$^{37,38}$.
Yet, precisely such an ability to ``compute'' in an arbitrary superposition 
would be necessary for an observer to be able to ``perceive'' an arbitrary 
quantum state. Moreover, in the external Universe only these observables which 
are resistant to decoherence and which correspond to ``pointer states'' are
worth recording: Records are valuable because they allow for predictions, and
resistance to decoherence is a precondition to predictability.$^{17,20}$

It is instructive to look at the evolving quantum system through the eyes of
an effectively classical, but ultimately quantum observer. In what follows, 
we shall focus on the memory, which we shall assume consists of a large
number of two-state systems (``memory cells"). We shall assume that the
effective classicality is a consequence of decoherence -- memory cells are 
immersed in an appropriate environment, which does not perturb selected pointer 
states, but destroys superpositions between them. Information processing would 
proceed through appropriate (i. e., {\tt c-not} like) interactions between 
such memory cells. There is no fundamental distinction between this memory
device and a classical computer.

The measurement is initiated with an (again {\tt c-not} like) coupling between 
a subset of initially ``empty'' memory cells and an outside system ${\cal S}$. 
As a result of such an interaction, a one-to-one correspondence between the 
states $|s_i>$ of the monitored quantum system ${\cal S}$ and the record state 
$|r_i>$ of a subset of memory ${\cal M}$ cells shall be set up. Decoherence will
play a role already during measurements. Thus; (a) unless the observer stores 
the information in the pointer states of memory, and; (b) unless he measures 
einselected pointer states of the system, a rapid (decoherence timescale) loss 
of information will ensue. In a simplest bit-by-bit {\tt c-not} like example, 
Eq. (4.1), a perfect correlation of two sets of orthogonal states 
$\{|\tilde 0>, |\tilde 1>\}$ -- Schmidt basis states which may in general
differ from the stable pointer states $\{ |0>, |1>\}$ for both the system and 
the memory (subscripts ${\cal S}$ and ${\cal M}$, respectively) would lead from 
the entangled state:
$$ |\psi_{\cal S M }> \ = \ a|\tilde 0>_{\cal S} |\tilde 0>_{\cal M} ~ 
+ ~ b|\tilde 1>_{\cal S} |\tilde 1>_{\cal M} \ ,  \eqno(7.1)$$
to a density matrix diagonal in the common pointer basis, given by the products
of the respective pointer states of both the system and the memory: 
$$\rho_{\cal S M} \ =\ p_{00}|00><00| + p_{11}|11><11|+ p_{01}|01><01|+ p_{10}|10><10|\ .
\eqno(7.2)$$
Here $|01><01| = |0>_{\cal S}|1>_{\cal M}<0|_{\cal S}<1|_{\cal M}$, etc., and;
$$p_{00}\ =\ |a<0|\tilde 0>_{\cal S}<0|\tilde 0>_{\cal M}~
+~b<0|\tilde 1>_{\cal S}<0|\tilde 1>_{\cal M}|^2\ , \eqno(7.3{\rm a}) $$
$$p_{01}\ =\ |a<0|\tilde 0>_s<1|\tilde 0>_m~+~b<0|\tilde 1>_s<1|\tilde 1>_m|^2\ ,... \eqno(7.3{\rm b}) $$
with the analogous formulae holding for the other probabilities.

The loss of purity and the simultaneous loss of information is caused by the
decoherence which is eliminating entanglement (as it should, to bring about 
semblance of classicality) but which is exacerbated by the mismatch between the 
Schmidt basis in the post-entanglement $|\psi_{\cal SM}>$, and the pointer
states of ${\cal S}$ and ${\cal M}$. It is reflected in the presence of the 
``error terms", such as $|01><01|$ in $\rho_{\cal SM}$, Eq. (7.2). They appear 
with probabilities $p_{01},~p_{10}$, which tend to zero as the pointer states 
align in both the system (control) and in the memory (target). Generalization of
this {\tt c-not} example to a more complicated case of many states of the system
and corresponding states of memory is as straightforward conceptually as it is 
notationally cumbersome. We shall leave it to the imagination of the reader. 

In this simple bit-by-bit case$^{11}$ as well as in the more general measurement
situations the information can be lost to the environment through ``leaks'' 
located at; (A) the memory, and ; (B), the system. In case (A), the loss 
of information is associated with disentaglement 
and reduction of the wavepacket, and is the price paid for the classicality of 
the records$^{39}$. When the system is isolated and this is the only reason for 
the information loss, the record states will be mutually orthogonal. Each of 
them will correspond to a unique (but not necessarily pure or orthogonal$^{11}$)
{\it conditional} state of the system;
$$ \rho_{{\cal S}|r_i} \ = <r_i|\rho_{\cal SM}|r_i>/Tr<r_i|\rho_{\cal SM}|r_i>
\ , \eqno(7.4{\rm a})$$
which is implied by the record $|r_i>$. When $\rho_{{\cal S}|r_i} = |s_i><s_i|$ 
(the measurement is exhaustive) ``implication" is readily defined through the 
use of the conditional probability:
$$ p(s_i|r_i)=p(s_i,r_i)/p(r_i)=1 \ . \eqno(7.4{\rm b})$$ 
This is the ``collapse''. In case (B) the system leaks information, 
as its state will decohere when it was initially entangled with the memory in 
a Schmidt basis misaligned with the system pointer states, or because it 
dynamically evolves into such a misaligned state. 

Decoherence in memory (A) is associated with the reduction of the state vector. 
Decoherence in the system (B) will be principally responsible for the increase 
of entropy in an evolving, open quantum system. Minimizing information loss in 
case (B) is the motivation behind the predictability sieve$^{20,26,27}$: 
A wise selection of what is measured will allow for the optimal preservation of 
the correlation with the state of the system. Optimal choice of the measured 
observable is obviously desirable, as it provides initial conditions 
most useful for making predictions. 

We have seen previously how chaotic dynamics spreads regular initial states
into the Schr\"odinger cat - like superpositions. Moreover, and as a result of
the environment - induced decoherence, the von Neumann entropy of the system 
will increase at a rate given by the sum of the Lyapunov exponents, Eq. (6.4). 
How will these two effects be reflected in the memory of our observer?

When the system is isolated, any measurement which distinguishes between
a complete set of states (and corresponds to a complete set of observables 
$\hat O$) will be equally good from the point of view of the observer. Such a
measurement carried out at a time $t_0$ results in a record state $|r_i(t_0)>$ 
of some initial state which could be in principle mixed as in Eq. (7.4a),
although we shall assume for simplicity that it is a pure $|s_i(t_0)>$:  
$$ \rho_{\cal S M} = \sum_i p_{ii} |r_i(t_0)>|s_i(t_0)><r_i(t_0)|<s_i(t_0)| \ .
\eqno(7.5)$$
Hence, at least in principle, an observer who has a record $r_i$ should be able 
to perform a measurement of a complete set of observables 
($\hat O(t) = U(t-t_0) \hat O U^{-1}(t-t_0)$, where $U(t)$ is a unitary 
evolution operator, related in an obvious manner to the evolution operator
of the system) which at any later time $t$ 
have a predictable outcome $|s_i(t)> = U(t-t_0)|s_i(t_0)>$.

A sequence of records of such  a succession of measurements of $\hat O(t)$ 
occurring at subsequent instants is predictable -- the corresponding set 
of records $R_i = \{r_i(t_0), r_i(t_1),...\} = \{r_i^{(0)}, r_i^{(1)},...\}$ 
is {\it algorithmically simple}. That is, the whole sequence can be generated 
from one of the records (say, $r_i^{(0)}$) by a program containing 
the Hamiltonian of the system and a sequence of time intervals between the 
consecutive measurements. The size of such a minimal program in bits will 
be typically much less than the size of the raw, uncompressed $R_i$ itself.

The {\it algorithmic information content}$^{40-43}$ $K({\tt s})$ of a binary 
sequence ${\tt s}$ is a measure which can be used to characterize the 
predictability of the data set. It is defined as the size, in bits, of the 
smallest program $p^*({\tt s})$ which can generate the string ${\tt s}$ on 
universal classical\footnote{**}{Note that it does not matter for the definition of the
algorithmic information content whether the computer is quantum or classical. 
This is because a quantum computer can be in principle simulated on the 
classical computer. The fact that for some operations quantum computer may be 
exponentially more efficient is irrelevant -- neither the time used nor the 
memory required matter for the definition of $K({\tt s})$. It is nevertheless 
intriguing to enquire about a {\it quantum algorithmic information content} of 
a state defined by the least number of {\it qubits} needed to produce a certain quantum state.} computer:
$$ K({\tt s}) = |p^*({\tt s})| \ . $$
Here ``$|...|$'' is size in bits. (It should be clear from the context below 
when ``$|...|$'' means `size in bits' or `absolute value of a complex number'.)
We shall not discuss here the technical details of this definition. 
Suffice it to say that -- using these ideas -- it is possible to develop a theory
of information content which is formally analogous to the Shannon information 
theory, but which measures algorithmic randomness (algorithmic information 
content, Kolmogorov complexity -- these terms are often used interchangeably) 
of specific records, and makes no appeal to probabilities$^{44,45}$. 

In the example of the record sequences $R_i$ discussed above:
$$ K(R_i) \leq  K(\hat H) + K(t_0,t_1,...) + K(r(t_0)) \ . \eqno(7.6) $$
That is, the size of a program required to predict the whole sequence $R_i$
of measurement records can be reduced to the Hamiltonian, the sequence of the
measurement instants, and the initial condition. This disparity between the
size of the record $R_i$ and the size of the minimal program necessary to 
generate it $K(R_i)$  -- the fact that $|R_i| \gg K(R_i)$ -- is the defining 
feature of the algorithmic simplicity.

A sequence of repeated measurements of an energy eigenstate is a common 
(if somewhat trivial and atypical) example of such a perfectly predictable 
``evolution'' resulting in an algorithmically simple record set $R_i$. In this 
case $U(t-t_0)$ is an identity (up to an overall phase), and, hence 
$K(R_i) = K(r_i^{(0)})$, give or take a few bits. In more typical examples of 
non-trivial evolution $|s(t)>$ would really change with time, and the inequality
(7.6) would be saturated. But, in any case, for a perfectly isolated quantum 
system, observer can in principle accomplish $K(R_i) \ll |R_i|$ 
by electing to measure observables that can be predicted on the basis of his 
initial measurement record. 

This last proviso -- the ability to measure evolved initial condition -- is not
at all trivial. In many situations -- including chaotic quantum systems -- the 
states $|s_i(t)>$ will be wildly exotic ``Schr\"odinger cat - like'' 
superpositions. And a typical observer will have  his disposal a rather limited 
set of measurable observables, often resulting in 
a single fixed set of potential outcome states $|\sigma_i>$. When such an 
observer attempts to re-measure an evolving quantum system he ``knows to be''
in a state $|s_i(t)>$ (which may have evolved from one of the $|\sigma_i>$'s), 
his memory will be first entangled with the state of the system, and then -- 
after a memory decoherence time -- it will suffer ``reduction'' to end up in 
a correlated mixture of different (memory) pointer states:
$$ \rho_{{\cal S M}|r_i^(t)}^{(t+1)}  \  = \ |r_i^{(t)}><r_i^{(t)}| \sum_j|<\sigma_j|s_i(t+1)>|^2|\sigma_j~r_j^{(t+1)}><\sigma_j~r_j^{(t+1)}|  \ .\eqno(7.7)$$
Here we have pulled out a common factor -- observer's record of the previous
state of the system -- and renormalized the conditional $\rho_{{\cal SM}|r_i}$
at the time $t+1$ right after a new measurement. This illustrates a single step 
in the reduction of the state vector caused by the coupling between the 
``apparatus pointer'' -- records in the memory of the observer -- and 
the environment in the case of exhaustive measurement.

A succession of such reductions will be represented by a sequence of branching 
records and a similarly co-diverging Hilbert space ``trajectories'' of the 
system punctuated by the recorded states. Two record sequences which originate 
from the same initial condition may initially coincide for a while, but will 
eventually diverge. Seen from the outside, the memory of the observer will be 
described by a density matrix $\rho_{\cal M}$, a mixture of all the possible 
records $\{ R \}$, each weighted with the probability $p(R)$. The associated
von Neumann entropy will be:
$$ {\cal H_{\cal M}} \ = \ - Tr \rho_{\cal M} \lg \rho_{\cal M} \ = \ -\sum_R
p(R) \lg p(R) \ ,
\eqno(7.8)$$
where $\lg = \log_2$, so that ${\cal H}$ is measured in bits. The entropy of the
records will increase at the rate set by the nature of measurements as well as
by the dynamics of the system, which will decide the distribution of the record
probabilities. For instance, in a chaotic system monitored on a time scale of 
the order of the dynamical time or less, entropy will typically increase at the 
rate given by the Lyapunov exponents, even in the absence of the environment. 
However, that rate is really set by the relation between the states which can be
predicted on the basis of the past records, and the states the observer actually
measures. Thus, when isolated, even a chaotic system can be made perfectly 
predictable providing that the measured observable is energy. Conversely, 
a trivial evolution -- a rotating spin ${1 \over 2}$ -- can produce entropy 
when measured at time intervals equal to a quarter of its period of rotation.
Indeed, even in the absence of any evolution of the system, alternating measurements
of spin ${1 \over 2}$ along any two orthogonal axes will result in the same
linear rate of entropy increase (linearity is characteristic of chaos in usual 
dynamical context -- see Eq. (6.4)). In all of these examples the increase of 
entropy is subjective -- it depends on the choices of measurements carried out 
by the observer. Yet, it is real: Unpredictable sequences of records which 
cannot be compressed to a single initial condition will remain as evidence.
The information is ultimately lost to the environment through the process (A).

We note that even when the entropy of the system is bounded (as would be 
the case for the spin ${ 1 \over 2}$) the entropy of the records can grow
without a bound -- the memory is idealized here as an infinite, initially
blank tape of a Turing machine.  

A whole sequence of such measurements viewed from the vantage point of the 
observer will result in a specific set of records -- a particular record state 
$|R> = |r_1 r_2 r_3...>$ in his memory, and on the diagonal of $\rho_{\cal M}$. 
Each specific $|R>$ will appear with the probability $p(R)$ given by the product
of consecutive conditional probabilities computed from the states corresponding 
to individual record sequences in a manner illustrated in Eq. (7.7). 
This probability cannot increase with addition of new events to the ``recorded 
history''. It can however stay constant when the initial record sets the initial
condition which allows for prediction of all subsequent records with certainty. 
Typically, however, the evolution will not be so perfectly predictable. 
Moreover, it can be shown that optimal record compression strategies (which have
to be applied to the set of all record sequences $\{R\}$) will have $K(R) 
\simeq - \log_2 p(R)$. 

The examples we have discussed above illustrate loss of information caused by 
decoherence of the records -- the ``collapse" process (A). The observer could in
principle avoid such ``subjective'' decreases of predictability by measuring 
only these observables he can predict with certainty. Unfortunately, in practice
this is not an option -- the observers can only measure certain observables of 
the system (and the corresponding set of their eigenstates), while the 
environment invariably imposes the same set of pointer states in the memory of 
the observer, thus eliminating entanglement. The increase of entropy is an 
inevitable consequence of that last step, which guarantees classicality of 
records. In absence of decoherence quantum records could be used by an isolated
quantum observer to enhance predictability.\footnote{***}{In simple examples (such 
as the quantum {\tt c-not} repeated many times) one can see how retention of 
entanglement is reflected in an increase of predictability, resulting in 
recurrences of the initial state. However, in this purely quantum context 
concepts such as predictability are harder to define, so these remarks should
be taken with a grain of salt.}

The most realistic (and most intriguing) case involves an evolving, decohering
quantum system which is also occasionally monitored by the observer. Let us 
first suppose that the observer is skillful, always able to ``match'' the basis
which diagonalizes the instantaneous conditional density matrix of the system 
based on his records with the choice of measurements. Then the measurement 
itself will not lead to a collapse -- rather, it will ``reveal'' to the observer
the system in one of the ``preexisting'' einselected pointer states.
The observers record $R$ will simply reflect the evolution of the system under 
the influence of the environment. Its algorithmic information content will grow,
on the average, with the increase of the entropy of the system which is now 
caused by the process (B) -- the decohering influence of the environment on the 
system. Hence, the average increase of the algorithmic information $\Delta K(R)$
per measurement will be given by the {\it conditional} algorithmic information 
defined as the number of ``extra bits'' required to extend the old record:
$$ \Delta <K(R)> \ = \ <K(R_{n+1}|R_n)> \ = \ \Delta {\cal H}_{\cal M} 
\ \simeq \ \Delta t <\dot {\cal H}_{\cal S}> \ . \eqno(7.9)$$
For a chaotic quantum system $<\dot{\cal H_{\cal S}}>$ is, on the average (over 
the phase space or -- equivalently -- over the ensemble of observer records 
weighted by their probabilities) given by the sum of positive Lyapunov 
exponents, Eq. (6.4).

We have thus arrived at the ``observer's own'' version of the second law: The 
observer will be affected by the unpredictability directly, as the algorithmic 
randomness of his record $K(R)$ of the enfolding history of the system will 
increase with the number of measurements. Moreover, that entropy increase will 
be now more objective in that observer will not be able to prevent it solely by
adjusting the measured observables of the system. But what if the observer does 
not measure? In that case the entropy will initially increase at essentially 
the same rate, although the unpredictability will be reflected in the ignorance 
of the observer,  measured by ${\cal H}(\rho_{\cal S})$, which will grow
with time, rather than in the clutter of the record of the unfolding history. 
Both effects can be taken into account simultaneously by introducing 
{\it physical entropy}$^{46}$, a measure  of disorder which includes both the 
statistical entropy ${\cal H}(\rho_{{\cal S}|R})$ attributed to the system by 
the observer who has the record $R$, and the algorithmic entropy $K(R)$:
$$ {\cal Z}(R) \ = \ K(R) \ + \ {\cal H}(\rho_{{\cal S}|R}) \ . \eqno(7.10) $$
This quantity combines a measure of the observers cost of storing the record 
$R$ and the ignorance remaining -- the fact that the system is described by 
$\rho_{{\cal S}|R}$ conditioned upon this specific record -- in spite of the 
past measurements. 

Physical entropy was introduced to remove the illusion that the observer can 
violate the second law by measuring: It shows that what happens in measurements 
is a replacement of ignorance (and simplicity) with information (and clutter), 
and that the sum of these two contributions for equilibrium cases yields 
a conservation law involving ${\cal Z}$$^{46}$. As such, it helps put the 
second law  on a firmer footing by introducing a quantity which the intelligent 
being -- attempting to act, for instance, in the Maxwell's demon capacity 
-- can apply to itself$^{46,47}$.

Here the role of ${\cal Z}$ is to connect the irreducible size of the record
$R$ (which increases suddenly upon measurement, but is constant in between) 
with the von Neumann entropy of a system which ``shrinks" upon a measurement.
Physical entropy -- when averaged over an ensemble of all records consistent
with the initial state of the system with the appropriate weights -- does not 
change when measurements carried out by the observers are reversible, i.e., 
adjusted so that they commute with the density matrix describing the system. 
However, imperfect measurements will add to the entropy production. Thus, when 
the system is kept far from equilibrium by the measurements of the observer:
$$ <{\cal Z}(R)> \ = \ Tr_{\cal M} [ \rho_{\cal M}
(K(R) ~+ ~ {\cal H}(\rho_{{\cal M}|R})) ]
\ \ge \ {\cal H}_{\cal S} \ . \eqno(7.11)$$

$K(R)$ for a specific sequence of skillful measurements carried out in time 
intervals small compared to the time it takes for the system to reach 
equilibrium will increase at the rate set by the dynamics. This is ultimately
due to $\dot {\cal H}_{{\cal S}|R}$ of an evolving system in between observer's 
measurements (but in the presence of environmental monitoring resulting 
in decoherence). Increase at the same dynamically determined rate is then a good
description of the behavior of ${\cal Z}(R)$ (their sum), the physical entropy
relevant for a particular observer with a specific measurement record. There 
can be of course ``atypical'' branches, characterized by a relative simplicity 
of the records. However, on the average (that is, when contributions of 
individual branches labelled by distinct records are added with the appropriate 
weights, Eq. (7.11)), the increase of physical entropy far from equilibrium
will be almost exactly equal to the increase of entropy caused by 
the combination of dynamics and decoherence. Thus, $<\dot {\cal Z}> ~ \simeq ~ <\dot {\cal H_{\cal S}}>$, which in 
a chaotic system will be give by Eq (6.4), while in a regular system 
$<\dot {\cal Z}> ~ \sim ~ 1/t$, and the physical entropy will increase slowly 
(logarithmically) with time.

These estimates are contingent upon an assumption of a skillful observer -- 
an observer who is able to measure the system using pointer observables -- that 
is, without causing additional decoherence (and consequent entropy increases) 
by his measurements. This may not be a realistic assumption, especially when 
the observer is attempting to resolve the density matrix of the system into the 
individual pure states. Indeed, especially in the case of such high-resolution 
measurements, the optimal basis may be somewhat dependent on the outcomes of the
preceding measurements -- on the observer's record$^{48,20}$. Thus, the increase
of physical entropy ${\cal Z}$ will most likely be more rapid than the increase 
of the von Neumann entropy of the same system in the absence of measurements.

There is another reason why the physical entropy may increase not only faster, 
but to larger values than the ordinary von Neumann entropy: The size of the 
observer's record will continue to increase beyond the time after which 
the system left on its own would have reached equilibrium. Thus, even after the 
density matrix of the system averaged over all of the measurement records has 
reached its equilibrium distribution, the conditional density matrix implied by 
a particular record $R$ (and perceived by an observer in possession of that record)
may be very far from equilibrium, and the system will continue to increase its 
physical entropy at a rate no less than $<\dot {\cal H}(\rho_{{\cal S}|R})>$. 
For example, in a chaotic quantum system physical entropy will not saturate at 
the equilibrium value: Rather, the record size can continue increasing forever 
(or at least until the memory shall run out of empty space) at the rate given 
by Eq. (6.4).

We have already noted that the increase of resolution will typically increase 
the rate of entropy production. This is because quantum measurements involve
reduction of the state vector -- a process associated with the entropy increase.
So, an observer may gain short-term predictability by making a high-resolution 
measurement, but unless he is a skillful observer, he will lose a certain 
amount of information through process (B), at the instant of measurement. 
This additional source of entropy increase will obviously depend on the 
frequency and the resolution of measurements. The extra penalty will be 
usually less when the resolution is lower. 

The above discussion illustrates two complementary points of view of the entropy
increase. ``Outsiders'' viewpoint limits the information loss to the process (B)
-- outsider does not monitor the system, and lets his ignorance grow (but 
his memory remains uncluttered). ``Insider'' measures the system repeatedly.
He knows more, but uses up his memory to record data which may soon become 
obsolete (i. e., in a chaotic system obsolescence sets in on a Lyapunov time). 
Decoherence has allowed us to carry out this quasi-classical analysis in 
a completely quantum setting.

It is too early to claim that all the issues arising in the context of the 
transition from quantum to classical have been settled with the help of 
decoherence. Decoherence and einselection are, however, rapidly becoming
a part of the standard lore$^{49-52}$. Where expected, they deliver classical
states, and -- as we have seen above -- guard against violations of the
correspondence principle. The answers which emerge may not be to everyone's 
liking, and do not really discriminate between the Copenhagen Interpretation
and the Many Worlds approach. Rather, they fit within either mold, providing
effectively the missing elements -- delineating the quantum-classical border
postulated by Bohr (decoherence time fast or slow compared to the dynamical 
timescales on the two sides of the ``border''), and supplying the scheme for 
defining distinct branches required by Everett (overlap of the branches 
is eliminated by decoherence).

\bigskip
\noindent{\bf Acknowledgments}

John Archibald Wheeler has a legendary ability to anticipate and stimulate
exciting developments in physics. I had the good luck to be at the University 
of Texas at the right time, where he convinced me (and a few others) that 
quantum physics should be understood (rather than just applied as 
a calculational tool). I would also like to thank Juan Pablo Paz, who
has been my close collaborator in the study of the competition between 
the quantum superposition principle and decoherence in the chaotic setting. 
I am grateful to Chris Jarzynski, Raymond Laflamme, and Michael Nielsen
for useful comments  on the manuscript. Last not least, I am grateful to 
the Nobel Foundation for a stimulating, enjoyable, and most memorable meeting.

\vfill
\eject

\noindent{\bf References}

1. Poincar\'e, H., {\it Les Methodes Nouvelles de la M\'echanique C\'eleste},
(Gauthier-Villars, Paris, 1892).

2. Laskar, J., {\it Nature} {\bf 338}, 237 (1989).

3. Sussman, G. J., and Wisdom, J., {\it Science} {\bf 257}, 56-62 (1992).

4. Wisdom, J., Peale, S. J., and Maignard, F. {\it Icarus} {\bf 58}, 137 (1984).

5. Wisdom, J., {\it Icarus} {\bf 63}, 272 (1985).

6. Berman, G. P., and Zaslavsky, G. M., {\it Physica} (Amsterdam) {\bf 91A},
450 (1978).

7. Berry, M. V., and Balzas N. L., {\it J. Phys. A} {\bf 12}, 625 (1979).

8. Zurek, W. H., and Paz, J. P., {\it Phys. Rev. Lett.} {\bf 72}, 2508-2511
(1994); {\it ibid.} {\bf 75}, 351 (1995).

9. These results were announced already some time ago, but appear to be still
somewhat controversial; see selected papers in Casati, G., and Chrikov, B., 
{\it Quantum Chaos} (Cambridge University Press, Cambridge, 1995).

10. Schr\"odinger, E., {\it Naturwissenschaften} {\bf 23}, pp. 807-812, 823-828,
844-849 (1935); English translation in Wheeler, J. A. and Zurek, W. H., eds.,
{\it Quantum Theory and Measurement}, pp. 152-167 (Princeton University Press,
Princeton, NJ, 1983).

11. Zurek, W. H., {\it Phys. Rev.} {\bf D 24}, 1516-1524 (1981).

12. Zurek, W. H., {\it Phys. Rev.} {\bf D 26}, 1862-1880 (1982).

13. Joos, E., and Zeh, H. D., {\it Zeits. Phys.} {\bf B 59}, 229 (1985).
 
14. Zurek, W. H., pp. 145-149 in {\it Frontiers of Nonequilibrium Statistical 
Mechanics}, G. T. Moore and M. O. Scully, eds. (Plenum, New York, 1986).

15. Milburn, G. J., and Holmes, C. A., {\it Phys. Rev. Lett.} {\bf 56}, 
2237-2240 (1986).

16. Haake, F., and Walls, D. F., in {\it Quantum Optics IV}, J. D. Harvey and 
D. F. Walls, eds. (Springer, Berlin, 1986).

17. Zurek, W. H., {\it Physics Today} {\bf 44}, 36-46 (1991).

18. Gell-Mann, M., and Hartle, J. B., {\it Phys. Rev.} {\bf D47}, 3345-3382 (1993)

19. Albrecht, A., {\it Phys. Rev.} {\bf D 48}, 3768 (1993).

20. Zurek, W. H., {\it Progr. Theor. Phys.} {\bf 89}, 281-302 (1993).

21. von Neumann, J., ``Measurement and reversibility'' 
and ``The measuring process'', chapters V and VI if 
{\it Mathematische Grundlagen der Quantenmechanik}, (Springer, Berlin, 1932);
English translation by R. T. Beyer 
{\it Mathematical Foundations of Quantum Mechanics}, (Princeton Univ. Press, Princeton, 1955).

22. Einstein, A., Podolsky, B., and Rosen, N., 
{\it Phys. Rev} {\bf 47}, 777-780 (1935).

23. Bohm, D., {\it Quantum Theory}, 
(Prentice-Hall, Engelwood Cliffs, 1951).

24. Bell, J. S., {\it Physics} {\bf 1}, 195-200 (1964).

25. Aspect, A., Dalibard, J., and Roger, G., {\it Phys. Rev. Lett} {\bf 49},
1804-1807 (1982).

26. Zurek, W. H., Habib, S., and Paz, J. P., {\it Phys. Rev. Lett} {\bf 70},
1187-1190 (1993); Anglin, J. R., and Zurek, W. H., {\it Phys Rev.} {\bf D53},
7327-7335 (1996).

27. Gallis, M. R., {\it Phys. Rev.} {\bf A53}, 655-660 (1996); Tegmark, M., and
Shapiro, H. S., {\it Phys. Rev.} {\bf E50}, 2538-2547 (1994).

28. Lindblad, G., {\it Comm. Math. Phys.} {\bf 40}, 119-130 (1976).

29. Caldeira, A. O., and Leggett, A. J., {\it Physica} {\bf 121A}, 587-616 
(1983); {\it Phys. Rev.} {\bf A31}, 1059 (1985).

30. Unruh, W. G., and Zurek, W. H., {\it Phys. Rev.} {\bf D40}, 1071-1094 (1989).

31. Brune, M., Hagley, E., Dreyer, J., Ma\^itre, X., Maali, A., Wunderlich, C., Raimond, J-M., and Haroche, S., {\it Phys. Rev. Lett.} {\bf 77}, 4887-4890 (1996).

32. Monroe, C., Meekhof, D. M., King, B. E., and Wineland, D. J., {\it Science},
{\bf 272}, 1131-1136 (1996).

33. Poyatos, J. F., Cirac, J. I., and Zoller, P., {\it Phys. Rev. Lett.} {\bf 77}, 4728-4731 (1996).

34. Habib, S., Shizume, K., and Zurek, W. H., {\it Phys. Rev. Lett}, in press.

35. Ott, E., Antonsen, T. M., and Hanson, J, {\it Phys. Rev. Lett.} {\bf 35},
2187-2190 (1984); Dittrich, T., and Graham, R., {\it Phys. Rev.} {\bf A42}, 
4647-4660 (1990), and references therein.

36. H. D. Zeh, {\it The Physical Basis of the Direction of Time}, {Springer,
Berlin, 1989).

37. Landauer, R., {\it Phil. Trans. R. Soc.} {\bf 353} 367 (1995); also, in  
{\it Proc. of the Drexel-4 Symposium on Quantum Nonintegrability: Quantum -- 
Classical Correspondence}, D. H. Feng and B.-L. Hu, eds. (World Scientific, 
Singapore, 1998); W. G. Unruh, {\it Phys. Rev} {\bf A51}, 992 (1995)
Chuang, I. L., Laflamme, R., Shor, P., and Zurek, W. H., {\it Science},
{\bf 270}, 1633-1635 (1995)

38. C. H. Bennett, {\it Physics Today} {\bf 48}, No. 10 (1995).

39. Zurek, W. H., ``Information transfer in quantum measurements'', pp. 87-116
in {\it Quantum Optics, Experimental Gravity, and the Measurement Theory},
P. Meystre and M. O. Scully, eds. (Plenum, New York, 1983).

40. Solomonoff, R. J., {\it Inform. Control} {\bf 7}, 1-22 and 224-254 (1964).

41. Kolmogorov, A. N., {\it Problems of Information Transmission} {\bf 1}, 4-7
(1965).

42. Chaitin, G. J., {\it J. Assoc. Comp. Mach.} {\bf 16} 145-159 (1966).

43. Martin-L\"of, P., ``Algorithmen and zuf\"allige Folgen'', lecture notes,
University of Erlangen, 1966.

44. Li, M., and Vitanyi, P., {\it An Introduction to Kolmogorov Complexity and
its Applications} (Springer, New York, 1993).

45. Cover, T. M., and Thomas, J. A., {\it Elements of Information Theory},
(Wiley, New York, 1991).

46. Zurek, W. H., {\it Phys. Rev.} {\bf A40} 4731-4751 (1989); {\it Nature}
{\bf 341} 119-124 (1989).

47. Caves, C. M., pp. 47-89 in {\it Physical Origins of Time Asymmetry}, 
J. J. Halliwell, J. P\'erez-Mercader, and W. H. Zurek, eds. (Cambridge Univ.
Press, Cambridge, 1994); C. M. Caves and R. Schack, {\it Complexity} {\bf 3},
46-57 (1997).

48. Paz, J. P., and Zurek, W. H., {\it Phys. Rev.} {\bf D48}, 2728-2738 (1993).

49. Gell-Mann, M., and Hartle, J. B., in {\it Complexity, Entropy, and the
Physics of Information}, W. H. Zurek, ed. (Addison-Wesley, Reading, 1990).

50. Omn\`es, R., {\it Rev. Mod. Phys.}  {\bf 64}, 339-382 (1992), and {\it The
Interpretation of Quantum Mechanics}, (Princeton, 1994).

51. Griffiths, R. B., {\it Phys. Rev.} {\bf A54}, 2759-2774, (1996).

52. Giulini, D., Joos, E., Kiefer, C., Kupsch, J., Stamatescu, I.-O., and 
Zeh, H. D., {\it Decoherence and the Appearance of a Classical World in Quantum 
Theory}, (Springer, Berlin, 1996).
\vfill
\eject
\noindent{\bf Figures}
\medskip
\noindent Fig. 1. Information transfer in measurements and in decoherence.

a) Controlled not ({\tt c-not}) as an elementary bit-by-bit measurement. Its
action is described by the ``truth table'' according to which the state of the
target bit (apparatus memory in the quantum measurement vocabulary) is 
``flipped'' when the control bit (measured system) is $|1>$ and untouched 
when it is $|0>$ (Eq. (4.1)). This can be accomplished by the unitary 
Schr\"odinger evolution (see, e. g., Ref. 11).

b) Decoherence process ``caricatured'' by means of {\tt c-not}s. Pointer state 
of the apparatus (and, formerly, target bit in the pre-measurement, Fig. 1a)
now acts as a control in the continuous monitoring by the {\tt c-not}s of the
environment. This continuous monitoring process is symbolically ``discretized''
here into a sequence of {\tt c-not}s, with the state of the environment 
assuming the role of the multi-bit target. Monitored observable of the apparatus
-- its pointer observable -- is in the end no longer entangled with the system, 
but the classical correlation remains. Decoherence is associated with the 
transfer of information about the to-be-classical observables to 
the environment. Classically, such information transfer is of no consequence.
In quantum physics it is absolutely crucial, as it is responsible for the 
effective classicality of certain quantum observables.

c) Noise is a process in which a pointer observable of the apparatus is 
perturbed by the environment. Noise differs from the purely quantum decoherence 
-- now the environment acts as a control, and the {\tt c-not}s which represent 
it act in the direction opposite to the decoherence {\tt c-not}s. Usually, both 
decoherence and noise are present. Preferred pointer observables and the 
associated pointer states are selected so that the noise is minimized.

\vfill\eject

\epsfbox{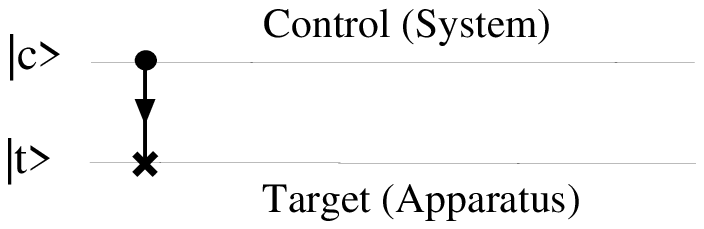}
\vskip 1truein

\epsfbox{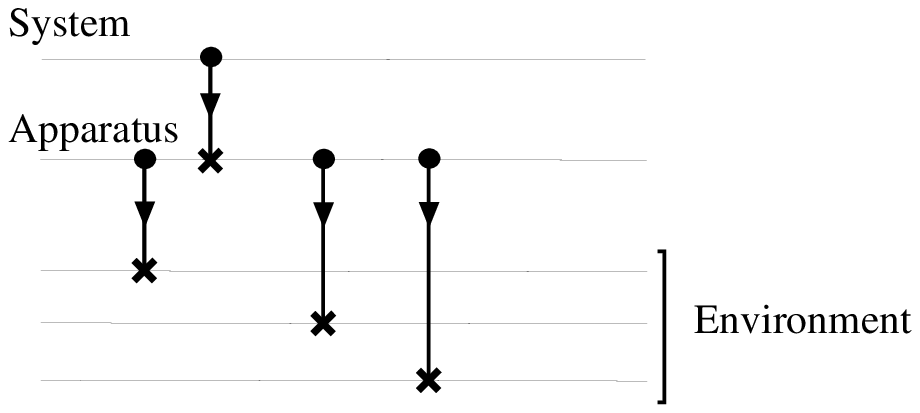}
\vskip 1truein

\epsfbox{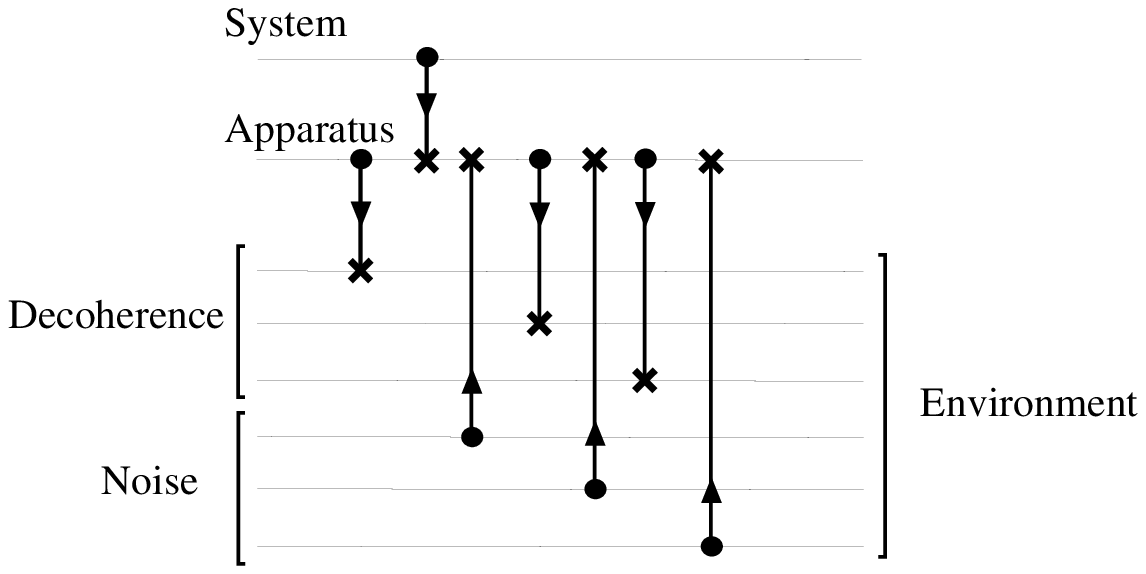}


\end